\documentclass[aps,groupedaddress,superscriptaddress,amsmath,amssymb,pra,twocolumn,longbibliography]{revtex4-2}
\usepackage{bm}
\usepackage{hyperref}
\usepackage{float}
\usepackage{graphicx}
\usepackage{verbatim}
\usepackage{bbm}
\usepackage{braket}
\usepackage{qcircuit}
\usepackage[normalem]{ulem}
\hypersetup{
    colorlinks=true,
    linkcolor=blue,
    filecolor=magenta,      
    urlcolor=cyan,
}
\usepackage{color}
\usepackage[utf8]{inputenc}

\def\llangle{\langle \!\langle}
\def\rrangle{\rangle \!\rangle}
\newcommand{\prm}[1]{^{( #1 )}}
\newcommand{\Gr}[1]{\langle\!\langle #1\big\rvert}

\def\so{\mathrm{so}}
\def\qb{{\bf q}}
\def\rb{{\bf r}}
\def\bn{{\bf n}}
\def\ep{\varepsilon}
\def\ka{\varkappa}
\def\ku{\ket{\uparrow}}

\def\kd{\ket{\downarrow}}

\begin{document}
\title{Spin dynamics in quantum dots on liquid helium}
\author{M. I. Dykman}
\affiliation{Department of Physics and Astronomy, Michigan State University, East Lansing, MI 48824, USA}
\author{Ofek Asban}
\affiliation{Department of Physics and Astronomy, Michigan State University, East Lansing, MI 48824, USA}
%\altaffiliation{These authors contributed equally to this work}
%
\author{Qianfan Chen}
\affiliation{Center for Nanoscale Materials, Argonne National Laboratory, Argonne, Illinois 60439, USA}
%\altaffiliation{These authors contributed equally to this work}
%
\author{Dafei Jin}
\affiliation{Department of Physics and Astronomy, University of Notre Dame, Notre Dame, IN 46556, USA}
\author{S. A. Lyon}
\affiliation{Department of Electrical and Computer Engineering, Princeton University, Princeton, NJ 08544, USA}

\begin{abstract} 
Liquid He-4 is free from magnetic defects, making it an ideal substrate for electrons with long-lived spin states. Such states can serve as qubit states. Here we consider the spin states of electrons electrostatically localized in quantum dots on a helium surface. Efficient gate operations in this system require spin-orbit coupling. It can be created by a nonuniform magnetic field from a current-carrying wire, can be turned on and off, and allows one to obtain large electro-dipole moment and comparatively fast coupling of spins in neighboring dots. Of central importance is to understand the spin decay due to the spin-orbit coupling. We establish the leading mechanism of such decay and show that the decay is sufficiently slow to enable high-fidelity single- and two-qubit gate operations.
\end{abstract}

\date{\today}

\maketitle

\section{Introduction}
\label{sec:Intro}

This paper is a contribution to the PRB Collection in honor of Emmanuel Rashba. It expands on Rashba's results on spin dynamics, and in particular the possibility of the electric dipole spin resonance where a resonant transition between spin states is excited by the electric field \cite{Rashba1960,Pekar1965}, see Ref.~\cite{Rashba2018} for a review.  Among numerous applications of these results of the immediate relevance is  spin dynamics in quantum dots. Such dynamics underlies the operation of semiconductor-based qubits \cite{Burkard2021}, a topic of significant current interest and an area of rapid progress, cf. \cite{Corrigan2021,Noiri2022,Harvey-Collard2022,Philips2022} and references therein. 

Along with qubits based on the electron states in semiconductor quantum dots \cite{Loss1998} there were proposed qubits based on the electron states in quantum dots on the helium surface \cite{Platzman1999}. This latter proposal relied on using orbital states of electron motion normal to the helium surface (a charge qubit). It was later extended to using in-plane orbital states \cite{Schuster2010}. Whereas quantized orbital states of motion normal to the surface have been known for a long time \cite{Grimes1976a}, quantized in-plane orbital states have been observed on helium only recently \cite{Koolstra2019a}. Quantized orbital states with a long coherence time were recently observed for electrons floating above solid neon, a system in many ways similar to electrons floating above liquid helium \cite{Zhou2022}. 

An alternative idea is to develop spin-state qubits  based on  electrons in quantum dots on helium \cite{Lyon2006} and on solid neon \cite{Chen2022a}. 
There is a major difference between spin-based qubits in semiconductors and on the helium surface. The electrodes that form a quantum dot on helium have to be submerged beneath the surface by a distance $\gtrsim 0.1~\mu$m to avoid strong coupling of the electrons to the vibrational excitations in helium \cite{Andrei1997}. Therefore it is hard to make a multiple-dot structure that would rely on the controlled interdot tunneling and the exchange electron coupling.  However, as we discuss, there are other ways to couple electron spins in different quantum dots.

\begin{figure}
\includegraphics[scale=0.8]{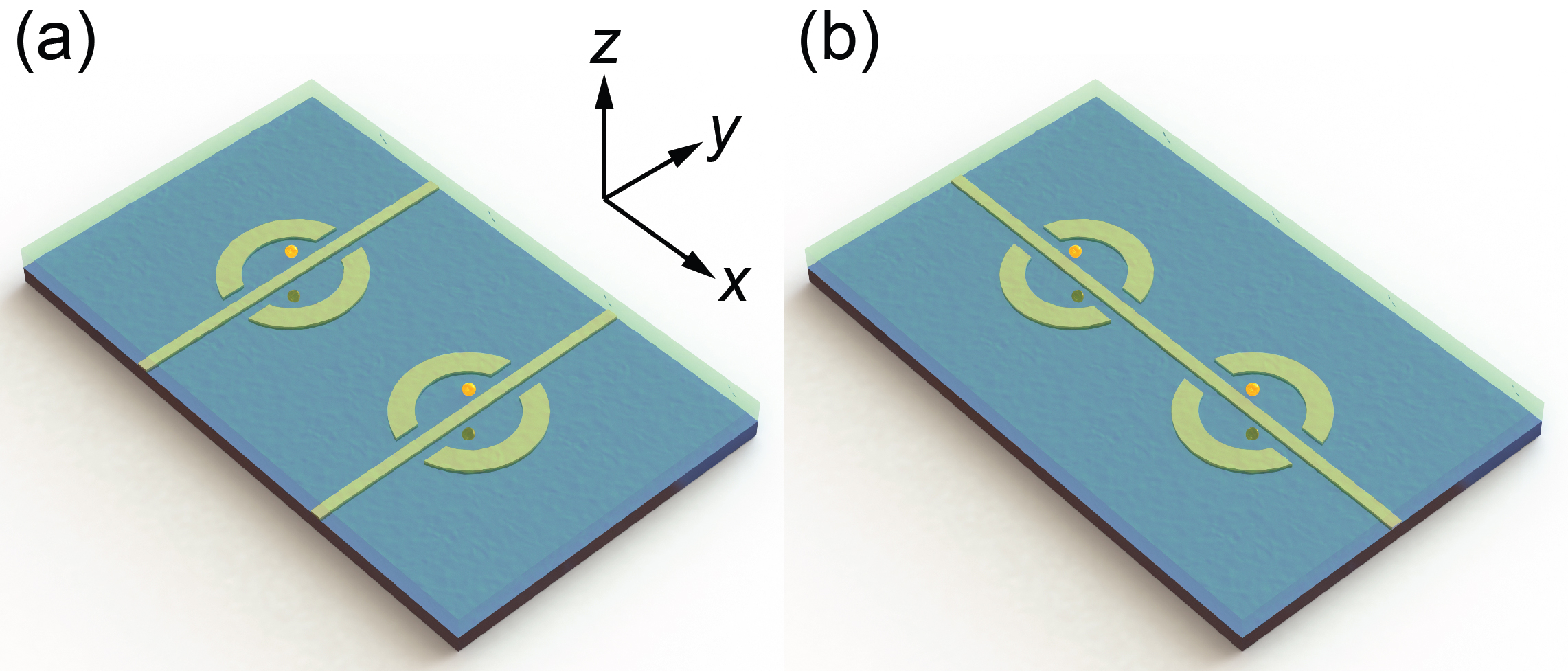}
%\hfill \includegraphics[scale=0.3]{Capture1.png}
\caption{Sketches of single-electron quantum dots on the surface of liquid helium. The confining electrodes are submerged into the helium to depth $0.2 - 0.5~\mu$m. The electron vibrations within the dot have typical frequencies of 6~GHz. A permanent magnetic field $\lesssim 0.5$~T is applied parallel to the surface. Superconducting wires running across the dots produce a nonuniform magnetic field that is turned on and off by a dc current $\sim 2 - 5$~mA through the wires. (a) The current can be turned on and off in each dot separately. (b) The current runs through several dots, and the spin-orbit coupling is controlled by tuning the intradot vibration frequency close to resonance with the Larmor frequency.}
\label{fig:sketch_dot}
\end{figure}

In this paper we study spin dynamics in a quantum dot on the helium surface.  The scheme of the quantum dot is sketched in Fig.~\ref{fig:sketch_dot}. The confining potential is created electrostatically. A strong magnetic field applied parallel to the surface leads to quantization of the spin states, which become the qubit states. The transitions between the states are controlled by pulses of a resonant microwave field. The coupling to this field can be significantly increased by the spin-orbit coupling, in the spirit of the electric dipole spin resonance. For helium, such coupling can be provided by a spatially nonuniform magnetic field from a wire above the dot \cite{Schuster2010}. Strong spin-photon coupling induced by a nonuniform magnetic field has been discussed and observed in quantum dots in semiconductors \cite{Samkharadze2018,Mi2018}. An advantageous feature of electrons on helium is that it should be possible to switch this coupling on an off. The effect is particularly significant if the spin transition frequency is close to the frequency of one of the orbital intrawell transitions.

Along with the increased coupling to the electromagnetic field, the induced spin-orbit coupling leads to relaxation of the electron spin. We argue that the major mechanism of spin relaxation is the decay of the excited spin state into the close in energy orbital state. On helium, the decay is induced by the coupling to surface capillary waves, ripplons.  Ripplons are soft excitations. The coupling to low-frequency ripplons involved in a small energy transfer can become strong even where the rate of the ripplon-induced scattering of unconfined electrons is still very small. In this case the spin relaxation rate is determined by processes with participation of many ripplons. As we show, for electrons on helium, by varying the temperature and the coupling parameters one can switch between weak- and strong-coupling limits. 

The qubit relaxation rate is important not only for the implementation  of single spin operations with microwave pulses, but also for operations on coupled spins, which correspond to two-qubit gate operations. Such operations can be performed between remote spins again using the aforementioned induced spin-orbit coupling \cite{Zhang2012}. The spin coupling in this case is a consequence of the Coulomb coupling between the electrons. The analysis of the resulting spin-spin coupling requires some care and is also discussed in the paper. 

In Sec.~\ref{sec:He_dots}  we describe the model of a single-electron quantum dot with coupled orbital vibrations of the electron and its spin. Section~\ref{sec:qubit_cplng} gives the theory of  the Coulomb-force mediated spin-spin coupling in different dots. Section~\ref{sec:relaxation} is the central section of the paper. It describes the mechanism of spin relaxation due to the spin-orbit coupling and the coupling of the orbital motion to capillary waves on the helium surface and shows where this coupling becomes strong and how it can be reduced to make a spin-based qubit. In Sec.~\ref{sec:estimates} we provide numerical estimates of the parameters. Section~\ref{sec:conclusions} provides concluding remarks. 

%The Rashba spin-orbit coupling of electrons on helium is extremely weak. Its effect on the electron dynamics and the weak ``intrinsic'' mechanisms of electron spin relaxation on helium, which do not come from the applied nonuniform magnetic field, will be discussed in a separate paper.

%%%%%%%%%%%%%%%%%%%%%%%%%%%%%%%%%%%%%%%%%%
%%%%%%%%%%%%%%%%%%%%%%%%%%%%%%%%%%%%%%%%%%%%%%

\section{Quantum dots on helium}
\label{sec:He_dots}

We will consider the geometry where the helium surface is in the $xy$-plane. The electrons float above the surface and are pressed against it by the electric field $E_\perp$, which is pointing along the $z$ axis. Quantum dots on the surface are created by electrodes submerged into the helium to depth $\lesssim 0.5~\mu$m \cite{Dykman2003a,Schuster2010,Koolstra2019a}. The low-lying orbital single-electron intradot states are weakly non-equidistant states of two vibrational modes. We will choose the coordinates of the normal mode to be pointing along the $x$ and $y$ axes. For brevity, we will call the modes $x$ and $y$ modes, respectively. The mode eigenfrequencies  $\omega_x$ and $\omega_y$ 
are controlled by the electrode potential. Of interest for the experiment is the range where 
$\omega_x/2\pi, \omega_y/2\pi$ 
are of the order of a few gigahertz; in particular, the frequencies used in the experiment \cite{Koolstra2019a} lied in this range. It is advantageous as it the range of frequencies of the superconducting microwave cavities used to excite inter-state transitions and measure qubit states in other implementations of condensed-matter based qubits. The frequencies $\omega_x$ and $\omega_y$ 
can be close to each other or different depending on the electrode geometry. 

The spin states of the localized electron are controlled by the magnetic field $B_x$, which in our model is applied along the $x$-axis. The Larmor frequency $\omega_L = 2\mu_BB_x/\hbar $ is also assumed to be in the few gigahertz range (here $\mu_B$ is the Bohr magneton and we have approximated the $g$ factor by 2). For a typical field $B_x=0.2$~T, which can be used in superconducting resonators,  $\omega_L/2\pi \approx 5.6$~GHz.  

We consider using the ground $\ket{\uparrow}$ and excited $\ket{\downarrow}$ spin states as the qubit states $\ket{0}$ and $\ket{1}$, respectively. They are eigenstates of the spin operator $s_x$,
\[\ku = 2^{-1/2}\left(\begin{array}{cc}1\\1\end{array}\right), \quad \kd = 2^{-1/2}\left(\begin{array}{cc}1\\-1\end{array}\right).\]
The temperature is assumed to be low, so that thermal excitation of the vibrational intradot states and the spin states can be disregarded. We will consider the setting in which the $y$-mode is not excited during the qubit operation and remains in its ground state. 

For a few low-lying orbital states the anharmonicity of the intradot vibrations can be disregarded, and then the Hamiltonian of the electron in the dot takes the form
\begin{align}
\label{eq:H0}
&H_0=\sum_{i=x,y}\hbar\omega_i a_i^\dagger a_i - \hbar\omega_L s_x,
\end{align}
where $a_i$ and $a_i^\dagger$ ($i=x,y$) are the ladder operators of the $x$ and $y$ vibrational modes. In what follows we will consider resonant behavior of the vibrations and the spin. To do this, we note that, for a system with the Hamiltonian $H_0$, in the Heisenberg representation the relevant operators oscillate in time as $a_i(t)\propto \exp(-i\omega_it)$ and $\tilde s_\pm(t) \propto \exp(\mp i \omega_Lt)$, where $\tilde s_{\pm} = s_y\pm i s_z$ (the commutation relations for these spin operators are $[\tilde s_\pm,s_x] = \mp \tilde s_\pm, [\tilde s_+,\tilde s_-]= 2s_x$).

The spin-orbit coupling comes from the coordinate-dependent magnetic field from a wire above the quantum dot, see Fig.~\ref{fig:sketch_dot}. We assume that the wire goes through the center of the dot along the direction $y$ in which one of the electron vibrational modes is polarized. The $z$-component of the field changes sign at the center of the dot, so that, if we count the $x$-coordinate of the electron off from the center of the dot, to the leading order the coupling to the $z$-field component is 
\begin{align}
\label{eq:induced_cplng}
&H_\so = \lambda_\so (a_x + a_x^\dagger)s_z, \nonumber\\
& \lambda_\so =2 l_x\mu_B\partial_xB_z, \quad l_x = (\hbar/2m_e\omega_x)^{1/2}
\end{align}
($m_e$ is the electron mass). We note that the field from the wire also changes the field $B_x$. In the analysis of the effect of the wire field we will assume that this change $\delta B_x$ is incorporated into the value of $B_x$. However one should keep in mind that the current through the wire is supposed to be turned on and off, so the field $B_x$ becomes time-dependent, which leads to accumulation of the qubit phase $\propto\int dt \delta B_x(t)$. The spatial nonuniformity of the field $\delta B_x$ is small for $l_x$ small compared to the distance to the wire.

\subsection{Electric dipole moment of the spin transition}
\label{subsec:dipole_moment}

If the Larmor frequency $\omega_L$ is close to $\omega_x$, the spin-orbit coupling (\ref{eq:induced_cplng}) is close to resonant. It strongly increases the oscillator strength of the spin-flip transition induced by the electromagnetic field making resonant spin transitions electrodipolar, reminiscent of the electrodipolar transitions described in Ref.~\cite{Rashba2018}.  A qualitative argument provided in Ref.~\cite{Schuster2010} suggested that, if the electron Hamiltonian in the dot is $H_0+H_\mathrm{so}$, the coupling of the spin transition to a resonant electric field is described by the Hamiltonian 
 \begin{align}
 \label{eq:e_dipolar}
 H_\mathrm{ed} =  -d_s s_zE_x, \quad d_s = -2\frac{e \mu_B\partial_xB_z}{m_e(\omega_L^2 - \omega_x^2)}.
 \end{align}
The parameter $d_s$ is the electric dipole moment of the resonant spin transition. It sharply increases if  the orbital and spin states are close to resonance, i.e. if $\omega_x$ is close to $\omega_L$, with $d_s\propto |\omega_L - \omega_x|^{-1}$.
 
A more rigorous analysis can be done by studying the electric conductivity of the system $\sigma_{xx}(\omega)$ in the frequency range $\omega\approx \omega_L$. The analysis  of $\sigma_{xx}(\omega)$ is given in Appendix~\ref{sec:conductivity}. The analysis is perturbative, it refers to the case where the coupling is comparatively weak, $|\lambda_\mathrm{so}/\hbar(\omega_L-\omega_x)|\ll 1$. For stronger coupling one should take into account the repulsion of the levels of the $x$-polarized vibrational mode and the spin. 

The conductivity $\sigma_{xx}(\omega)$ displays a peak at $\omega_L$. The amplitude of this peak is determined by the dipole moment $d_s$.  The corresponding electro-dipolar absorption can be much larger than the relativistically-small magneto-dipolar absorption.

%%%%%%%%%%%%%%%%%%%%%%%%%%%%%%%%%%%%%%%%%%%%%%%%%%%%%%%%
%%%%%%%%%%%%%%%%%%%%%%%%%%%%%%%%%%%%%%%%%%%%%%%%%%%%%%%%

\section{Spin-spin coupling}
\label{sec:qubit_cplng}

The spin-orbit coupling induced by the nonuniform field $B_z$ leads also to the coupling of the spins in different quantum dot. It is mediated by the Coulomb coupling between the electrons in different dots and by the intradot vibrations. %The  Coulomb coupling comes via the hybridization of the intradot vibrations.% and the second is the direct dipolar coupling.
%\subsection{Coupling via hybridized intradot vibrations.}
%\label{subsec:hybridized_cplng}
%
The coupling via intradot vibrations was considered by Zhang et al. \cite{Zhang2012}. We now provide a  derivation that gives a different result from Ref.~\cite{Zhang2012}.  It allows us to obtain what we believe is  a reliable estimate of the spin-spin coupling.

The electron spin and the electron vibrations in the same dot are coupled by the $B_z$ field. A key observation for describing the spin coupling in different dots is that the vibrations in different dots are hybridized by the Coulomb coupling. Then the spins in different dots become coupled to the same hybridized modes. This leads to the spin-spin coupling mediated by virtual exchange of vibrational excitations, similar to the spin coupling via microwave cavity photons in double quantum dots in semiconductors \cite{Harvey-Collard2022}. 

We consider first a geometry where the two dots are aligned along the $x$-axis and are at a distance $R$ from each other, with $R\gg l_{x}\prm{n}, l_{y}\prm{n}$, see Fig.~\ref{fig:sketch_dot}~(a). Here we use the superscript $n=1,2$ to enumerate the dots. In particular, $\lambda_\mathrm{so}\prm{n}$ is the spin-orbit coupling parameter (\ref{eq:induced_cplng}) for the dot $n$ and $\omega_{x,y}\prm{n}$ are the vibration frequencies in this dot; $l_{y}\prm{n}=(\hbar/2m_e\omega_{y}\prm{n})^{1/2}$. The strong field $B$ along the $x$-axis is the same for the both dots. However, the nonuniform field $B_z$ that couples the spins in each dot to the intradot vibrations can be different and the frequencies $\omega_x\prm{n}$ can depend on $n$ as well. 

The interdot coupling energy of the electrons $V_\mathrm{id}(|\rb\prm{1}-\rb\prm{2}|)$ can differ from $e^2/|\rb\prm{1} - \rb\prm{2}|$ because of screening. To the lowest order in the displacements $x\prm{1,2}$ from the equilibrium intradot position, the $x$-displacement-dependent term in the potential energy of the vibrations has the form
\begin{align}
\label{eq:x_potential}
&U(x\prm{1},x\prm{2}) \approx \frac{1}{2}m_e\sum_n \omega_x\prm{n}{}^2 x\prm{n}{}^2 +\frac{1}{2}m_e\omega_C^2(x\prm{1}-x\prm{2})^2,\nonumber\\
&\omega_C^2 =   m_e ^{-1}[d^2V_\mathrm{id}(r)/dr^2]_{r=R}. 
\end{align}
Here we have taken into account that the coupling potential $V_\mathrm{id}(r)$ falls off with the increasing $r$. The term $\propto (x\prm{1}-x\prm{2})^2$ in $U(x\prm{1},x\prm{2})$ describes the coupling of  the intradot vibrations. 

The potential (\ref{eq:x_potential}) can be diagonalized by a standard rotation
\begin{align}
\label{eq:diagonalization}
&x\prm{1}+ix\prm{2} = ( x_1+i x_2)e^{-i\alpha_{12}}, \nonumber\\
&\tan 2\alpha_{12} = 2\omega_C^2/\left(\omega_x\prm{1}{}^2 -\omega_x\prm{2}{}^2\right).
\end{align}
The same transformation has to be done for the $x$-components of the momenta to obtain normal modes.
The squared eigenfreqiencies $\Omega_{1,2}{}^2$ of the vibrations with the collective coordinates $x_{1,2}$ are, respectively,
\begin{align}
\label{eq:eigenfreqiencies}
&\Omega_k{}^2 = \frac{1}{2}(\omega_x\prm{1}{}^2 +\omega_x\prm{2}{}^2)  +\omega_C^2 
-\frac{(-1)^k}{2}\left[2\omega_C^2\sin(2\alpha_{12})\right.\nonumber\\
&\left. +(\omega_x\prm{1}{}^2-\omega_x\prm{2}{}^2)\cos(2\alpha_{12}) \right]\qquad (k=1,2).
\end{align}

As a result of the hybridization, each hybridized mode is coupled to both spins. If  $c_k$ and $c_k^\dagger$ are the lowering and raising operators of the hybridized modes, the coupling has the form
\begin{align}
\label{eq:coupled_prime}
H'_\mathrm{so}= \sum_{kn} \lambda_k\prm{n}\left( c_k + c_k^\dagger\right) s_z\prm{n},
\end{align}
where $k=1,2$ enumerates the hybridized modes and $n=1,2$ enumerates the dots, 
\begin{align}
\label{eq:coupled_lambda}
&\lambda_n\prm{n} =(\omega_x\prm{n}{}/\Omega_n)^{1/2}\lambda_\mathrm{so}\prm{n}\cos\alpha_{12},\nonumber\\
&\lambda_k\prm{n}= (-1)^k(\omega_x\prm{n}/\Omega_k)^{1/2}\lambda_\mathrm{so}\prm{n}\sin\alpha_{12}\; (n\neq k).
\end{align}

Coupling of the spins to the same mode leads to the spin-spin coupling. For the coupling to one orbital mode the effect is described in Appendix ~\ref{sec:Green}. Extending the result to the case of the coupling to two modes, we obtain for the exchange Hamiltonian $H_\mathrm{ex}$ 
\begin{align}
\label{eq:full_spin_coupling}
&H_\mathrm{ex} = g\tilde s_+\prm{1}\tilde s_-\prm{2} + \mathrm{H.c.}, 
\quad 
g= \frac{1}{4\hbar} \lambda_\mathrm{so}\prm{1}\lambda_\mathrm{so}\prm{2}\sin(2\alpha_{12})\nonumber\\
&\times \frac{(\omega_x\prm{1}\omega_x\prm{2})^{1/2}
(\Omega_2{}^2-\Omega_1{}^2)
%\omega_C^2\sin(2\alpha_{12})
}
{(\omega_L^2 - \Omega_1{}^2) (\omega_L^2 - \Omega_2{}^2)}.
\end{align}

From Eq.~(\ref{eq:full_spin_coupling}), the exchange coupling parameter is $g\propto \omega_C^2$ for  $\omega_C^2 \gg |\omega_x\prm{2}{}^2 - \omega_x\prm{1}{}^2|$ (this limit is reached for any $\omega_C$ if the intrawell frequencies $\omega_x\prm{1}$ and $\omega_x\prm{2}$ are close) and $|2\omega_L^2 -\omega_x\prm{1}{}^2 -\omega_x\prm{2}{}^2|\gg \omega_C^2$ . The relation $g\propto \omega_C^2$ also holds for a weak Coulomb coupling,  $\omega_C^2 \ll |\omega_x\prm{2}{}^2 - \omega_x\prm{1}{}^2|, |\omega_L^2-\omega_x\prm{1,2}{}^2|$. 

The dependence of $g$ on the frequency $\omega_C$ is modified in the important case where both frequencies $\omega_x\prm{1,2}$ are close to each other and to $\omega_L$. Here, for a small interdot distance, one can have $\omega_C^2 \gg |\omega_x\prm{2}{}^2 - \omega_x\prm{1}{}^2|$ and $\omega_C^2 \gg |\omega_L^2 - \bar\omega_x^2|$, where $\bar\omega_x^2=( \omega_x\prm{1}{}^2 +\omega_x\prm{2}{}^2)/2$. Then
\begin{align}
\label{eq:interdot_simplified}
g\approx \frac{1}{4\hbar} \lambda_\mathrm{so}\prm{1}\lambda_\mathrm{so}\prm{2} \bar\omega_x (\omega_L^2 - \bar\omega_x^2)^{-1}.
\end{align}
In this case $g\propto |\omega_L -\bar\omega_x|^{-1}$
sharply increases as $\omega_x$ approaches the Larmor frequency, but is practically independent of the interdot distance. 
We note that $\omega_L$ is shifted by the polaronic effect from the coupling to the intradot electron vibrational modes. This shift is discussed in Appendix~\ref{sec:Green}. 

The results immediately extend to the geometry where the wires in the dots are along the inter-dot direction, see Fig.~\ref{fig:sketch_dot}~(b), if the current in each dot can be switched on an off separately.  The orbital coupling energy in this case has the same form as in Eq.~(\ref{eq:x_potential}), except that $\omega_C^2$ has to be replaced with $-\omega_C^2 = -(m_eR)^{-1} |dV_\mathrm{id}/dr|_{r=R}$. Respectively, $\omega_C^2$ has to be replaced by $-\omega_C^2$ in Eqs.~(\ref{eq:diagonalization}) - (\ref{eq:interdot_simplified}). 

To conclude this section we note that it is tempting to think that, since the spins are associated with electric dipoles,  cf. Eq.~(\ref{eq:e_dipolar}), there should be a direct electrodipolar coupling between spins in different dots.  If the dipole moment operators in the dots 1 and 2 are $d_s\prm{1}s_z\prm{1}$ and $d_s\prm{2}s_z\prm{2}$, respectively, and screening can be disregarded, the energy of the dipole-dipole coupling is $-(d_s\prm{1} d_s\prm{2}/R^3)s_z\prm{1}s_z\prm{2} $. It leads to the coupling of the form Eq.~(\ref{eq:full_spin_coupling}), but the value of the coupling parameter $g$ is different from that in Eq.~(\ref{eq:full_spin_coupling}). However, the underlying argument is misleading. To describe the spin-spin coupling one has to consistently take into account the coupling of the spins to the respective intradot vibrations and the coupling between the vibrations in different dots.

%%%%%%%%%%%%%%%%%%%%%%%%%%%%%%%%%%%%%%%%%%%%%%%%%%%%
%%%%%%%%%%%%%%%%%%%%%%%%%%%%%%%%%%%%%%%%%%%%%%%%%%%%

\section{Spin relaxation due to the imposed spin-orbit coupling}
\label{sec:relaxation}

The orbital motion of the electron is coupled to vibrational modes in helium or neon. These modes are described by a two-dimensional (2D) wave vector $\qb$ and a branch number $\mu$; for example, $\mu$ can refer to ripplons or phonons in liquid helium or different acoustic branches in solid neon.  Their Hamiltonian $H_b$ and the coupling Hamiltonian $H_i$ are
\begin{align}
\label{eq:Hi}
&H_b=\hbar\sum_{\qb,\mu}\omega_{\qb\mu}b^\dagger_{\qb\mu}b_{\qb\mu},
\nonumber\\ 
& H_i=\sum_{\qb,\mu}V_{\qb\mu}e^{i\qb\rb} \left(b_{\qb\mu} + b^\dagger_{-\qb\mu}\right),
\end{align}
where $\omega_{\qb\mu}$ is the frequency and $b_{\qb\mu}$ is the annihilation operator of mode $(\qb,\mu)$, $\rb = (x,y)$ is the electron coordinate, and $V_{\qb\mu}$ is the parameter of the electron coupling to mode $(\qb,\mu)$. It is obtained by projecting the overall coupling energy onto the lowest state of electron motion normal to the surface. Parameters $V_{\qb\mu}$ are well-known for the coupling to ripplons and phonons on helium \cite{Dykman2003a,Schuster2010} and for the coupling to phonons on the neon surface \cite{Zhou2022,Chen2022a}.

It is a good approximation to disregard the direct coupling of ripplons and phonons to the spin of an electron floating above the surface. This is an important distinction from spin relaxation in  quantum dots in semiconductors \cite{Khaetskii2001,Golovach2004,Zwanenburg2013,Burkard2021}.

The spin-orbit coupling induced by the nonuniform magnetic field introduces the coupling of the spin not only  to the microwave electric field but also to ripplons and phonons. It thus leads to spin relaxation. For quantum computing applications, the relaxation time has to be long so that the spin does not have a chance to decay during the time when the field $B_z$ is on. Calculating the $B_z$-induced relaxation time is the goal of this section. We will concentrate on the spin relaxation of electrons on the helium surface, the extension to the electrons on the surface of solid neon is straightforward.

An important aspect of the electron dynamics on helium is that the major source of dissipation, the ripplons, are very soft excitations.  Transitions between  discrete electron states separated by a few gigahertz, in energy units, require emitting or absorbing two ripplons or a phonon \cite{Dykman2003a}. This leads to a long lifetime of the orbital states, which is estimated to be in the range of $0.1$~msec \cite{Dykman2003a,Schuster2010}. The idea of using spin states for quantum information is based on the assumption that the relaxation of the spin states will be still much longer. However, when an orbital state and a spin state are close in energy, single-ripplon processes can lead to transitions between them. This can significantly increase the spin relaxation rate. Moreover, the coupling to ripplons in the localized electron states may be strong. In this case the electron dynamics in quantum dots becomes similar to that in color centers, where the electron energy spectrum in the absence of the coupling to phonons is also discrete \cite{Pekar1950,Huang1950}. 

%%%%%%%%%%%%%%%%%%%%%%%%%%%%

\subsection{Polaron transformation}
\label{subsec:polaron}

To analyze the dynamics in the presence of the coupling to excitations with the energy smaller than the level spacing of the electron system it is convenient to write the electron Hamiltonian in the absence of spin-orbit coupling as

\begin{align}
\label{eq:H0_n_kappa}
&H_0=\sum_{\bn,\ka}\ep_{\bn, \ka}\ket{\bn\ka}\bra{\bn\ka},\nonumber\\
 &\ep_{\bn, \ka}=\hbar(\omega_xn_x+\omega_yn_y + \omega_L \ka).
\end{align}
Here $\bn\equiv (n_x,n_y)$ enumerates the excited states of $x$- and $y$ modes. Parameter $\ka$ enumerates the ground ($\ka=0$) and excited ($\ka=1$) spin states; $\ket{\bn\ka} = \ket{\bn}\ket{\ka}$. Equation (\ref{eq:H0_n_kappa}) differs from the expression (\ref{eq:H0}) for $H_0$ by the constant $-\hbar\omega_L/2$; this difference plays no role in the further analysis. In the $(\bn,\ka)$-representation the Hamiltonian of the coupling to the vibrational modes of helium reads
\begin{align}
\label{eq:Hi_n_kappa}
&H_i = \sum_{\qb,\mu,\bn,\bn'}V_{\qb\mu}^{\bn \bn'}\left(b_{\qb\mu} + b^\dagger_{-\qb\mu}\right) \ket{\bn}\bra{\bn'},\nonumber\\
&V_{\qb\mu}^{\bn \bn'} = V_{\qb\mu}\bra{\bn}\exp(i\qb\rb)\ket{\bn'}.
\end{align}
The operator $H_i$ is a unit operator in the spin space.

The effect of the coupling to the modes $(\qb,\mu)$ on the spin dynamics can be conveniently analyzed using the polaron unitary transformation
\begin{align}
\label{eq:polaron_transform}
U=\exp\left[\sum_{\qb,\mu,\bn}\left(V_{\qb\mu}^{\bn \bn}/\hbar\omega_{\qb\mu}\right)(b_{\qb\mu}-b_{-\qb \mu}^\dagger) \ket{\bn}\bra{\bn}\right] 
\end{align}
This transformation allows one to eliminate the diagonal in $\bn$ terms in the coupling Hamiltonian $H_i$, 
\begin{align*}
&U^\dagger (H_b+H_i)U =  H_b+\sum_\bn P_{\bn}\ket{\bn}\bra{\bn}\nonumber\\
& + U^\dagger\sum_{\qb,\mu,\bn\neq \bn'}V_{\qb\mu}^{\bn \bn'}\left(b_{\qb\mu} + b^\dagger_{-\qb\mu}\right) \ket{\bn}\bra{\bn'}U,  \nonumber\\
 &P_{\bn} = -\sum_{\qb\mu}|V_{\qb\mu}^{\bn\bn}|^2/\hbar\omega_{\qb\mu}.
\end{align*}
Here $P_{\bn}$ is the polaronic shift of the ${\bn}$th intradot orbital energy level. Where the level spacing in the quantum dot exceeds the typical frequencies $\omega_{\qb\mu}$ of the helium modes coupled to the electron, the off-diagonal in $\bn,\bn'$ terms in $H_i$ lead to shifts of the electron energy levels, which are smaller than $P_{\bn}$. These off-diagonal terms remain off-diagonal after the unitary transformation. In what follows they are disregarded. 

The transformation (\ref{eq:polaron_transform}) couples the spin states to modes $(\qb,\mu)$. We will consider this coupling assuming that the electron remains in the state $n_y=0$, since $y$-vibrations are uncoupled from the spin. Respectively, we will use the notation 
\[v_{\qb\mu}^{(m)} = (\hbar\omega_{\qb\mu})^{-1} [V_{\qb\mu}^{\bn\bn}]_{n_x = m,n_y=0}
.\]  
With these notations, the transformed Hamiltonian of the spin-orbit coupling (\ref{eq:induced_cplng}) becomes 
\begin{align}
\label{eq:spin_bath}
&U^\dagger H_\mathrm{so}U = \lambda_\mathrm{so}s_z\sum_m(m+1)^{1/2}\ket{m+1}\bra{m}\nonumber\\
&\times
\exp\left[\sum_{\qb,\mu}
\left(v_{\qb\mu}^{(m)} -v_{\qb\mu}^{(m+1)}\right)(b_{\qb\mu} - b^\dagger_{-\qb\mu})\right] + \mathrm{H.c.}.
\end{align}
where $\ket{m}\equiv \ket{n_x=m,n_y=0}$ and we used that $a_x^\dagger \equiv \sum_m (m+1)^{1/2}\ket{m+1}\bra{m}$.

%%%%%%%%%%%%%%%%%%%%%%%%%%%%%%%%%%%%%%%%%%%%%%

\subsection{Spin decay rate}
\label{subsec:decay_rate}

The spin-to-helium modes coupling described by Eq.~(\ref{eq:spin_bath}) leads to a decay of the excited spin state $\kd$. The involved process is a spin transition to the ground state, $\kd \to \ku$, which is  accompanied by the excitation of the orbital state $\ket{m}\to \ket{m+1}$ [we recall that $m$ here enumerates the states of $x$-polarized vibrations], with the energy difference going to helium modes.  

The physics strongly depends on the lifetime of the excited orbital state. We will assume that this time, even though it is long on the scale of the modern condensed-matter qubits, is still much shorter than the spin lifetime. Then, once the excitation has been transferred from the spin to the excited orbital state, it will not go back to the spin state. In this case the transition rate $W_s$ can be calculated keeping the leading-order term in $\lambda_\mathrm{so}$. We will assume that, initially, the $x$-mode is in the ground state. This means that, in terms of the states $\ket{\bn\ka}$, we are calculating the probability of the transition $\ket{n_x=n_y=0,\ka=1}\to \ket{n_x=1,n_y=0,\ka = 0}$. This probability has to be averaged over the states of the helium modes. A standard calculation gives   
\begin{align}
\label{eq:spin_rate}
&W_s =\frac{\lambda_\mathrm{so}^2}{2\hbar^2}\mathrm{Re}\,\int_0^\infty dt e^{i(\omega_L - \omega_x)t-\epsilon t}\exp[-w_\mathrm{pol}(t)],
\nonumber\\
&w_\mathrm{pol}(t) =\sum_{\qb,\mu}|\alpha_{\qb\mu}|^2\left[(2\bar n_{\qb\mu}+1)(1-\cos\omega_{\qb\mu} t) \right.
\nonumber\\
&\left. + i\sin\omega_{\qb\mu} t\right], \quad \bar n_{\qb\mu} \equiv \bar n(\omega_{\qb\mu}); \quad\epsilon\to +0.
\end{align}
Here  $\bar n(\omega) = [\exp(\hbar\omega/k_BT)-1]^{-1}$. The parameter $\alpha_{\qb\mu} = v_{\qb\mu}^{(0)} - v_{\qb\mu}^{(1)}$ is the difference of the coupling parameters in the ground and first excited states of the $x$-mode. Using the explicit form of the coupling parameters (\ref{eq:Hi_n_kappa}) and the wave functions of the electron intradot vibrations in the states involved in the transition, we obtain
\begin{align}
\label{eq:cplng_parameter}
\alpha_{\qb\mu} =(V_{\qb\mu}l_x^2 
q_x^2/\hbar\omega_{\qb\mu})\exp\left[-\frac{1}{2}(q_x^2l_x^2 + q_y^2l_y^2)\right].
\end{align}

The factor $\exp[-w_\mathrm{pol}(t)]$ is similar to the factor that determines the shape of the absorption lines of color centers in solids \cite{Pekar1950,Huang1950}. The spin relaxation rate is determined by the Fourier transform of this factor at the frequency $\omega_L-\omega_x$ given by the difference of the Larmor frequency and the frequency of the intradot vibrations coupled to the spin.

The general expression for $W_s$ simplifies in the limiting cases of weak and strong coupling of the intradot vibrations to the vibrations of liquid helium [note that Eq.~(\ref{eq:spin_rate}) refers to a  weak spin-orbit coupling; however, the coupling of the intradot vibrations to  the helium modes can be weak or strong]. For weak coupling we have for $W_s= W_s^\mathrm{weak}$,
\begin{align}
\label{eq:weak_cplng}
W_s^\mathrm{weak} =& \frac{\pi\lambda_\mathrm{so}^2}{2\hbar^2}\sum_{\qb,\mu}|\alpha_{\qb\mu}|^2
\left(\bar n_{\qb\mu} +\frac{1}{2}\pm\frac{1}{2}\right)\nonumber\\
&\times  \delta(\omega_L - \omega_x \mp\omega_{\qb\mu}).
\end{align}
Here the upper sign refers to the process where $\omega_L>\omega_x$ and the spin transition to the ground state is accompanied by an excitation of the intradot vibrations and emission of a helium vibration. The lower sign refers to the case $\omega_L <\omega_x$ where the spin transition requires activation by  helium vibrations.

For strong coupling the integral over time in Eq.~(\ref{eq:spin_rate}) can be calculated by steepest descent. This gives  $W_s =W_s^\mathrm{strong}$, where
\begin{align}
\label{eq:strong_cplng}
&W_s^\mathrm{strong}=\frac{\pi^{1/2}\lambda_\mathrm{so}^2}{2^{3/2}\hbar^2\gamma}\exp\left[-(\omega_L -\omega_x - P_\mathrm{strong})^2/2\gamma^2\right],\nonumber\\
&P_\mathrm{strong} = \sum_{\qb,\mu}|\alpha_{\qb\mu}|^2\omega_{\qb\mu},\nonumber\\
&\gamma=\left[\sum_{\qb,\mu}|\alpha_{\qb\mu}|^2(2\bar n_{\qb\mu}+1)\omega_{\qb\mu}^2\right]^{1/2}.
\end{align}
Equation (\ref{eq:strong_cplng}) applies provided 
\begin{align}
\label{eq:strong_cplng_condition}
\gamma^2\gg   \bar\omega_{\qb\mu}^2, 
 %  \bar\omega_{\qb\mu}|\omega_L -\omega_x - P_\mathrm{strong}|,
\end{align}
where $\bar\omega_{\qb\mu}$ is the typical frequency of the helium vibrations coupled to the electron.  The inequality (\ref{eq:strong_cplng_condition}) is sufficient where the relevant vibrations of helium are thermally excited, $\hbar \bar\omega_{\qb\mu}\ll k_BT$. The transition $\ket{\downarrow, n_x=0} \to \ket{\uparrow,n_x=1}$ is then accompanied by emission and absorption of a large number of vibrational helium excitations. If, on the other hand, $\hbar \bar\omega_{\qb\mu} \gg k_BT$, the strong-coupling limit applies for $\omega_L -\omega_x - P_\mathrm{strong}>0$. In this case the transition $\ket{\downarrow, n_x=0} \to \ket{\uparrow,n_x=1}$ is accompanied only by emission of helium excitations. 

The dependence of the switching rate on the frequency difference $\omega_L-\omega_x$ is Gaussian in the strong-coupling case, with a characteristic width $\gamma$. Parameter $P_\mathrm{strong}$ comes from the polaronic shift of the energy levels of the orbital states $\ket{n_x=0}$ and $\ket{n_x=1}$. It does not depend on temperature. In contrast, the width $\gamma$ is temperature-dependent.  It linearly increase with $T$ for $k_BT\gg \bar\omega_{\qb\mu}$.

We emphasize that $\alpha_{\qb\mu}$ are the parameters of the coupling of electron orbital motion to helium vibrations. They do not depend on the detuning of the spin transition frequency $\omega_L$ from the intradot electron frequency $\omega_x$. The strong coupling parameters $\gamma,P_\mathrm{strong}$ are also independent of $\omega_L-\omega_x$. However, the rate  $W_s$ depends on $\omega_L-\omega_x$ both directly, as seen from Eqs.~(\ref{eq:weak_cplng}) and (\ref{eq:strong_cplng}), and in terms of the spin-orbit coupling parameter $\lambda_\mathrm{so}$.

%%%%%%%%%%%%%%%%%%%%%%%%%%%%%%%%%%
%%%%%%%%%%%%%%%%%%%%%%%%%%%%%%%%%%%%%

\section{Numerical estimates for quantum dots on helium surface}
\label{sec:estimates}

In this section we provide estimates of the spin-spin coupling mediated by the induced spin-orbit coupling and of the spin decay rate. We consider a quantum dot with the intradot frequency and the Larmor frequency being $\omega_x\sim \omega_L \sim 2\pi\times 6$~GHz. We assume that the spin-orbit coupling is imposed by a wire parallel to the $y$-axis at a distance $0.5~\mu$m from the helium surface, and that the current in this wire is 2~mA. We set the difference between the Larmor frequency and the intradot vibration frequency  to be $|\omega_L-\omega_x|/2\pi = 5$~MHz, For these parameter values  the spin-orbit coupling constant $\lambda_\mathrm{so}/\hbar$ and the scaled electric dipole moment of the resonant spin transition  $|d_s/e|$ are  
\[\lambda_\mathrm{so}/\hbar \approx 1.1\times 10^7\,\mathrm{s}^{-1}, \quad |d_s/e| \approx 13.8\,\mathrm{nm}.\]

We estimate the exchange coupling between the spins in different dots assuming that the dots have the same frequencies $\omega_x$, that they are  3~$\mu$m apart, and that the Coulomb coupling between the electrons is not screened. Then from Eq.~(\ref{eq:full_spin_coupling}) the coupling parameter is
\[g/\hbar \approx 5.5\times 10^5~\mathrm{s}^{-1}.\] 
We note that, for the Coulomb coupling used in obtaining this estimate, the coupling frequency $\omega_C$ largely exceeds $|\omega_L -\omega_x|$, and therefore $g$ weakly depends on $\omega_C$.

The above estimates refer to the case where on controls spins in different dots independently by applying a nonuniform magnetic field independently to each dot, as in Fig.~\ref{fig:sketch_dot}~(a). In the setting shown in Fig.~\ref{fig:sketch_dot}~(b) a nonuniform magnetic field is applied globally to all dots at a time. The effect of the spin-orbit coupling can be controlled in this case by changing the intradot vibration frequency electrostatically. For example, if we want to suppress the spin orbit-coupling in a dot and, as before, we set $\omega_L\approx 2\pi\times 6$~GHz, we can detune $\omega_x$ to $2\pi\times 3$~GHz. Then the electric dipole moment becomes $|d_s/e| \approx 0.03$~nm and the interdot coupling parameter becomes $g\approx 9~\mathrm{s}^{-1}$. The spin relaxation  mechanism discussed earlier does not work, the spin decay rate becomes exceedingly small. On the whole, even though the control is less efficient than when the nonuniform magnetic field is turned on an off in each dot separately, the ``switching off'' the spin-orbit coupling electrostatically is still fairly efficient.

One of the most important estimates is that of the spin relaxation rate. We will consider it for scattering by ripplons. We will drop the subscript $\mu$ when discussing ripplon frequencies and the parameters of the coupling to ripplons. The characteristic ripplon frequency is 
\begin{align}
\label{eq:typical_freq}
\bar\omega \equiv \bar\omega_\qb = (\sigma_\mathrm{He}/\rho_\mathrm{He} l_x^3)^{1/2},
\end{align}
where $\sigma_\mathrm{He}$ and $\rho_\mathrm{He}$ are the surface tension and the density of liquid helium, and we use that the typical  wave number of the ripplons coupled to the electron is $q=l_x^{-1}$.  Numerically,  we have $\bar\omega \approx 2\times 10^8~\mathrm{s}^{-1}$. For the typical temperatures $T\gg 1.6$~mK used in the experiment we have $k_BT\gg \hbar\bar\omega_\qb$. 

The coupling to ripplons is determined by the electric field $E_\perp$ that presses the electrons against the helium surface and the polarization coupling that comes from the ripplon-induced modulation of the image potential \cite{Shikin1974a}. The coupling parameters $V_\qb$ are given by the sum of the contributions of these two mechanisms. Therefore $|\alpha_\qb|^2$, along with the sum of the corresponding contributions, has a cross term. However, this term is smaller than the other two terms and we will not evaluate it.

It is seen from Eqs.~(\ref{eq:weak_cplng})  and (\ref{eq:strong_cplng}) that of interest for the analysis of relaxation due to ripplon scattering is the function
\begin{align}
\label{eq:cplng_function}
\aleph(\omega)= \sum_\qb |\alpha_\qb|^2 (2\bar n_{\qb\mu} +1)\delta(\omega_\qb - \omega).
\end{align}  
To estimate $\aleph(\omega)$ we will assume that the localization lengths $l_x$ and $l_y$ are the same. Using the explicit form of the coupling parameters (cf. \cite{Dykman2003a}), we find that, for typical frequencies $\omega$ in the range $10^5 - 10^9\,\mathrm{s}^{-1}$, the values of $\aleph(\omega)$ for the $E_\perp$-induced coupling, $\aleph_{E_\perp}(\omega)$, is
\begin{align}
\label{eq:E_perp_cplng}
\aleph_{E_\perp}(\omega) = \frac{e^2E_\perp^2 k_BT}{8\pi\hbar^2}
\left(\sigma_\mathrm{He}\omega^{1/3}\bar\omega^{8/3}\right)^{-1}
e^{-(\omega/\bar\omega)^{4/3}},
\end{align}
whereas for the polarization-induced coupling, $\aleph_\mathrm{pol}(\omega)$, it is
\begin{align}
\label{eq:polarization_cplng}
\aleph_\mathrm{pol}(\omega)=\frac{\Lambda^2k_BT r_\mathrm{pol}^2}{32\pi\hbar^2}\frac{\rho_\mathrm{He}^{4/3}}{\sigma_\mathrm{He}^{7/3}}\,\frac{\omega^{7/3}}{\bar\omega^{8/3}}
e^{-(\omega/\bar\omega)^{4/3}}.
\end{align}
In Eq.~(\ref{eq:polarization_cplng}), $\Lambda = (\epsilon_\mathrm{He}-1)e^2/4(\epsilon_\mathrm{He}+1)$, where $\epsilon_\mathrm{He}$ is the helium dielectric constant; $r_\mathrm{pol} = (2/3)\log(\omega/\bar\omega) + \log(r_B/2l_x) + \gamma_E-(1/2)$, with $\gamma_E$ being the Euler constant and $r_B$ being close to the electron localization length normal to the helium surface, $r_B\approx 7.6~\mathrm{nm}$. 

%%%%%%%%%%%%%%%%%%%%%%%%%%%%%%%%%%%%%%%

\subsection{The strong-coupling parameters}
\label{subsec:strong_cplng}

The function $\aleph(\omega)$ immediately gives the strong-coupling parameter $\gamma$ and the polaronic shift $P_\mathrm{strong}$  due to the coupling to ripplons, which are defined in Eq.~(\ref{eq:strong_cplng}). For  $k_BT\gg \hbar\bar\omega$ 
\begin{align}
\label{eq:strong_cplng_aleph}
\gamma^2 = \int_0^\infty d\omega \,\omega^2 \aleph(\omega), \quad P_\mathrm{strong} = \hbar\gamma^2/2k_BT.
\end{align}
From Eq.~(\ref{eq:E_perp_cplng}), for the coupling due to the pressing field $E_\perp$, the strong coupling parameter is 
\[\gamma_{E_\perp} = (3e^2E_\perp^2k_BT/32\pi\hbar^2\sigma_\mathrm{He})^{1/2}.\]  
The expression for the contribution $\gamma_\mathrm{pol}$ to $\gamma$ due to the polarization interaction is cumbersome. For typical $l_x$ used earlier and for $r_B=7.6~\mathrm{nm}$ we have 
\[\gamma_\mathrm{pol} = [(3\Lambda^2k_BT/128\pi\hbar^2) \rho_\mathrm{He}^{4/3}\sigma_\mathrm{He}^{-7/3}\bar\omega^{8/3}C_\mathrm{pol}]^{1/2}\] 
with $C_\mathrm{pol}\approx 16$.

For $T=0.1$~K these expressions give $\gamma_\mathrm{pol}\approx 2.1\times 10^8~\mathrm{s}^{-1}$ and, for typical $E_\perp=300~\mathrm{V/cm}$, $ \gamma_{E_\perp}\approx 4.8\times 10^8~\mathrm{s}^{-1}$. Thus, $\gamma_{E_\perp}$ exceeds $\bar\omega$, whereas $\gamma_\mathrm{pol}$ is very close to $\bar\omega$, which indicates that the coupling to ripplons is not weak for such a temperature.

% Reaching weak coupling requires going to the temperature $\sim 10$~mK and reducing  the pressing field $E_\perp$.

%%%%%%%%%%%%%%%%%%%%%%%%%%%%%%%%%%%%%

\subsection{Estimate of the spin decay rate}
\label{subsec:decay_rate}

We recall the constraints on the spin decay rate $W_s$. First, it  must be small compared to the energy decay rate of the orbital state to which the energy is transferred  $W_\mathrm{orb}$.  Then there is an important constraint on the quantum-information side: the spin-decay rate should be small compared to the reciprocal duration of the gate operations. The latter means that, when the nonuniform magnetic field is on and gate operations are performed, the decoherence of the spin state is inessential.
The rate of the qubit (spin) swap gate, which is determined by the coupling parameter $g$ in Eq.~(\ref{eq:full_spin_coupling}), is $\propto \lambda_\mathrm{so}^2$ (if we assume the same spin-orbit coupling in the both dots). The qubit decay rate as given by (\ref{eq:spin_rate}) is also $\propto \lambda_\mathrm{so}^2$. Therefore it is the ripplon-dependent factor in the spin decay rate that must be adjusted to meet the low-decoherence condition. 

In the limit of  strong coupling to ripplons, from the condition $
W_s^\mathrm{strong}\sim \lambda_\mathrm{so}^2/\hbar^2\gamma \ll g/\hbar \sim \lambda_\mathrm{so}^2/\hbar^2 \Delta\omega$ we have $\gamma\gg \Delta\omega$, where $\Delta\omega = |\omega_L-\bar\omega_x|$ (in estimating $g$ we assumed that $\omega_C\gg \Delta\omega$). The  condition  $\gamma\gg \Delta\omega$ is incompatible with the above analysis of the electro-dipolar spin transition. Indeed, the parameter $\gamma$ determines the typical width of the orbital-transition spectral line at frequency $\omega_x$. This width is supposed to be much smaller than $\Delta\omega$, so that the spin-transition spectral line is well separated from the orbital line and the spin can be selectively excited by a resonant electric field at frequency $\omega_L$. %We note that the estimate  $\gamma\gg \Delta\omega$ refers to the arguably most relevant case where  $\Delta\omega$ is small compared to the frequency $\omega_C$ determined by the interdot Coulomb coupling.

Therefore of primary interest is the case of weak electron-ripplon coupling, $\gamma \ll \bar\omega$. In this case the spin decay rate is
\begin{align}
\label{eq:spin_weak_aleph}
W_s^\mathrm{weak} = (\pi\lambda_\mathrm{so}^2/4\hbar^2)\aleph(\Delta\omega),\quad \Delta\omega=|\omega_L-\omega_x|).
\end{align}
Here we have taken into account that $k_BT\gg \hbar\bar\omega, \Delta\omega$. In this expression, and in fact throughout the paper [except for Eq.~(\ref{eq:strong_cplng})], we assume that the orbital frequency $\omega_x$ has been renormalized to incorporate the frequency shift due to the coupling to ripplons.

The condition of the small effect of decoherence on the gate operation $W_s^\mathrm{weak}\ll |g|/\hbar$ and the condition of the spectral separation of the spin and orbital transitions $W_s^\mathrm{weak} \ll \Delta\omega$ are met for $\aleph(\Delta\omega)\ll \Delta\omega^{-1}$. The interdot coupling $g$ increases with the decreasing $\Delta\omega$. As seen from Eqs.~(\ref{eq:E_perp_cplng}) and (\ref{eq:polarization_cplng}), for typical  $\Delta\omega \ll \bar\omega$, the pressing-field induced contribution $\aleph_{E_\perp}(\Delta\omega)$  increases with the decreasing $\Delta\omega$ as $\Delta\omega^{-1/3}$, whereas the polarization contribution $\aleph_\mathrm{pol}(\Delta\omega)$ decreases as $\Delta\omega^{7/3}$.  

For the small detuning $\Delta\omega/2\pi = 5~\mathrm{MHz}$ that we consider and for $T=0.01$~K, we have  $\aleph_\mathrm{pol}(\Delta\omega), \aleph_{E_\perp}(\Delta\omega)\ll \Delta\omega^{-1}$. The polarization-coupling induced decay rate for weak coupling to ripplons is 
\[W_s^\mathrm{pol} \sim 0.6\times 10^3~\mathrm{s}^{-1}.\]
The $E_\perp$-induced decay rate is $\sim 6\times 10^5~\mathrm{s}^{-1}$ for the pressing field $E_\perp = 300$~V/cm often used in estimates, i.e., $W_s^{E_\perp}\sim g/\hbar$  . Therefore to meet the small-decoherence conditions it is necessary to significantly decrease the pressing field $E_\perp$ compared to 300~V/cm, which is a fairly routine procedure in the experiment.

The energy-decay rate of the orbital states on liquid helium is $W_\mathrm{orb}\sim 3\times 10^4~\mathrm{s}^{-1}$, for the typical intradot frequency $\omega_x\sim 2\pi\times 6~\mathrm{GHz}$ \cite{Dykman2003a,Schuster2010}. It is  larger than $W_s^\mathrm{pol}$ by a factor of 50. This justifies the considered model of spin decay. 

The probability of spin decay during the gate operation, if we estimate is as $\hbar W_s^\mathrm{pol}/g $, is $\sim 10^{-3}$. It can be further  reduced by reducing the difference $\Delta\omega$ of the Larmor frequency and the orbital intradot frequency. The constraint on $\Delta\omega$ from below is imposed by the linewidth of the intradot orbital transition. Because of the nonzero linewidth, it is possible to excite an orbital state by driving the electron at the Larmor frequency. However, the excitation probability is $\propto (W_\mathrm{orb}/\Delta\omega)^2$ and would remain negligibly small even if $\Delta\omega$ is further reduced by a factor of 10. 

The other constraint comes from the fact that the calculation has been based on the perturbation theory, which relies on the assumption that $\Delta\omega$ is much larger than $\lambda_\mathrm{so}/\hbar$, i.e., that the spin-orbit coupling is weak. The analysis of the range where the spin-orbit coupling is strong and the electron states should be described in terms of dressed states \cite{Orszag2016} is beyond the scope of this paper; this range is less interesting in terms of a spin qubit on helium. In the perturbation theory, the parameters $g$ and $W_s$ scale as $\lambda_\mathrm{so}^2$. Therefore the relations between the qubit parameters will not be compromised if the ratio $\lambda_\mathrm{so}/\Delta\omega$ is reduced, which can be done just by reducing the control field $B_z$.

\section{Conclusions}
\label{sec:conclusions}

The results of this paper show that the spin dynamics in quantum dots on liquid helium is very  different from the spin dynamics in quantum dots in semiconductors. The difference stems from several factors. In particular, there are no ohmic contacts to the electrons on helium. The quantum dots on helium are $\sim 0.2 - 0.5~\mu$m in diameter and the interdot distance is $\gtrsim 1-3~\mu$m, so that the excited low-lying orbital electron states are just intradot vibrational states, whereas the interdot tunneling plays no role. There are no fluctuating nuclear magnetic moments in $^4$He, and  the intrinsic spin-orbit coupling, including the Rashba coupling, is extremely weak  for electrons on helium. Therefore the spin relaxation time is exceedingly long. The leading sources of spin decoherence are magnetic fields from stray currents and vortices in the superconducting circuits, but those can be  efficiently suppressed by choosing proper materials, particularly given that the electrons are comparatively far away ($\sim 0.5~\mu$m) from the electrodes.  

A major advantageous feature of the system is the possibility to turn the spin-orbit coupling on and off by applying a spatially nonuniform magnetic field from a current-carrying wire. A Nb wire with a cross-section of $100\times 150$~nm can carry 5 mA, which should suffice to produce a strong enough field.  The spin-orbit coupling enables electro-dipolar transitions of the spin in a quantum dot, and thus strongly enhances the coupling of the spin to the electromagnetic mode in the cavity. This allows performing single-qubit gate operations on the spin in a conceivable time for a reasonable strength of the microwave field. The intradot spin-orbit coupling also enables coupling spins in different dots with each other much stronger than via a magnetic dipolar coupling. This opens a way of performing two-qubit gate operations at rate $\sim 5\times 10^5~\mathrm{s}^{-1}$.

In a broad parameter range the electric dipole moment of the spin transition and the interdot spin-spin coupling parameter scale as the inverse of the detuning $\Delta\omega= |\omega_L-\omega_x|$ of the Larmor frequency $\omega_L$ from the intradot vibration frequency $\omega_x$. The detuning can be controlled by the gate electrodes in a broad range. This allows one to  efficiently turn the spin-orbit coupling on and off electrostatically by changing $\omega_x$, in the presence of a spatially nonuniform magnetic field. Such arrangement is particularly important in the geometry where the nonunifrom field is turned globally for all quantum dots at a time, cf. Fig.~\ref{fig:sketch_dot}~(b).

The other side of the spin-orbit coupling is that it leads to decay of the spin states. Our estimates show that the decay rate can be made small, so that the spin coherence is preserved during a gate operation. A feature of the electrons on helium is that they are coupled most strongly to soft excitations, ripplons. For a typical wave number $q=2\times 10^5~\mathrm{cm}^{-1}$, the ripplon frequency is $\sim 1$~mK in the temperature units. The leading process of spin decay is a transition from the excited spin into the intradot vibrational state with the energy deficit $\hbar\Delta\omega$ going to or taken from a ripplon. It is this process that becomes slow for small (but not too small) $\Delta\omega$ provided the field $E_\perp$ that presses the electron against the helium surface is largely compensated.

Besides the spin physics, an interesting physics that can be explored with electrons in quantum dots on helium is related to the possibility to have strong coupling of the intradot vibrations to ripplons. The characteristic coupling energy can exceed the characteristic ripplon frequency. This makes intradot vibrations similar to the electron excitations in color centers in solids \cite{Pekar1950,Huang1950}. An important and  unique aspect of the electron dynamics on helium is that, in contrast to color centers, the coupling to ripplons can be controlled by varying the field $E_\perp$. Some features of these dynamics reminiscent of the dynamics of color centers were noticed before for the system, where electrons were not localized \cite{Chepelianskii2021}. An electron localized in a quantum dot is a much closer analog of a color center, which allows getting a new insight into the physics of color centers with a fully controlled coupling of the electron to the vibrations in the medium.

\acknowledgments

The research of M. I. D., O. A., and S. A. L. was supported in part by the Grant No. DE-SC0020136 funded by the U.S. Department of Energy, Office of Science. Q.C. and D. J. acknowledge support from the Argonne National Laboratory Directed Research and Development (LDRD).

%%%%%%%%%%%%%%%%%%%%%%%%%%%%%%%%%%%%%%%%%%%
%%%%%%%%%%%%%%%%%%%%%%%%%%%%%%%%%%%%%%

\appendix

\section{Resonant electro-dipolar spin conductivity}
\label{sec:conductivity}

It is convenient to express the conductivity $\sigma_{xx}(\omega)$ of an electron in a quantum dot in terms of a two-time Green function \cite{Bogolyubov1959,Zubarev1960}.  For operators $A$ and $B$, a two-time Green function is defined as
\begin{align}
\label{eq:green_defined}
\llangle A\big\vert B\rrangle_\omega = -i\int_0^\infty dt e^{i\omega t}\langle [A(t),B(0)]\rangle,
\end{align}
where $\langle\cdot\rangle$ stands for statistical averaging. The equation of motion for such Green function follows from the Heisenberg equation of motion for the operator $A$. For a system with the Hamiltonian $\mathcal{H}$ it reads
\begin{align}
\label{eq:Green_eom}
\omega \llangle A \big\vert B\rrangle_\omega = \hbar^{-1}\llangle [A,\mathcal{H}] \big\vert B\rrangle_\omega +\langle[A,B]\rangle .
\end{align}

The Kubo formula for the conductivity is
\begin{align}
\label{eq:Kubo}
\mathrm{Re}\,\sigma_{xx}(\omega) = (e^2\omega/\hbar)\,\mathrm{Im}\,\llangle x\big\vert x\rrangle_\omega
\end{align}
It is convenient to write $x=l_x(a_x+a_x^\dagger)$ and calculate the Green function for the ladder operators. If the Hamiltonian is $\mathcal{H} = H_0 +H_\mathrm{so}$, see Eqs.~(\ref{eq:H0}) and (\ref{eq:induced_cplng}), we have
\[(\omega-\omega_x)\llangle a_x\big\vert x\rrangle_\omega = (\lambda_\mathrm{so}/\hbar)\llangle s_z\big\vert x\rrangle + l_x.\]
Writing a similar equation for $\llangle a_x^\dagger\big\vert x\rrangle_\omega$, we obtain 
\begin{align}
\label{eq:first_xx_Green}
\llangle x\big\vert x\rrangle_\omega = \frac{2\lambda_\mathrm{so}l_x\omega_x}{\hbar(\omega^2-\omega_x^2)}\llangle s_z\big\vert x\rrangle_\omega +\frac{2l_x\omega_x}{\omega^2-\omega_x^2}.
\end{align}

To find the Green function  $\llangle s_z\big\vert x\rrangle_\omega$ we write the equation of motion (\ref{eq:Green_eom}) as
\[\omega \llangle A \big\vert B\rrangle_\omega = -\hbar^{-1}\llangle A\big\vert [B,\mathcal{H}] \rrangle_\omega +\langle[A,B]\rangle.\]
This equation follows from the Heisenberg equation of motion $dB/dt = -(i/\hbar)[B,\mathcal{H}]$.
Following the steps that led to Eq.~(\ref{eq:first_xx_Green}), we ultimately obtain
\begin{align}
\label{eq:conductivity_expl}
&\mathrm{Re}\, \sigma_{xx}(\omega) = (\omega/\hbar) \mathcal{D}^2\, \mathrm{Im}\, \llangle s_z\big\vert s_z\rrangle_\omega, 
\nonumber\\
&\mathcal{D} = 2e\mu_B\partial_x B_z/[m_e(\omega^2-\omega_x^2)]
\end{align}
In the frequency range $\omega\approx\omega_L$, which is of interest, we can replace $\omega$ with $\omega_L$ in the expression for $\mathcal {D}$, and then $\mathcal{D}s_z$ becomes equal to the dipole $d_s$ in Eq.~(\ref{eq:e_dipolar}). This means that the conductivity (\ref{eq:conductivity_expl}) is given by the electrical conductivity  of the dipole $d_s$, the latter being determined entirely by the electron spin.

%%%%%%%%%%%%%%%%%%%%%%%%%%%%%%%%%%%%
%%%%%%%%%%%%%%%%%%%%%%%%%%%%%%%%%%%

%\section{Exotic mechanisms of spin-orbit coupling for electrons on liquid helium and solid neon}
%\label{sec:exotic}

\section{Derivation of the spin-spin coupling}
\label{sec:Green}

Here we show how coupling of electron spins to the same vibrational mode leads to the spin-spin coupling. The mechanism is a standard exchange of virtual excitations, but it has its peculiar form in the problem of spins in quantum dots. Also, the Green function technique we use differs from what is typically done in the analysis of the coupling mediated by a microwave mode in a cavity. Besides the spin-spin coupling, it also describes the polaronic shift of the spin energies. 

We first show what the Green function technique gives where the sought interaction is already in place, i.e., we derive the equation of motion for the relevant Green function in this case. We then show how the same equation of motion  follows from the model where there is no direct interaction, but there is coupling to the same vibrational mode.  

The model of directly coupled spins in a magnetic field is described by the Hamiltonian $\mathcal{H}_{ee}$,
\begin{align}
\label{eq:toy_H}
&\mathcal{H}_{ee} = \mathcal{H}_0 + \mathcal{H}_d,\quad  \mathcal{H}_0= -\hbar \omega_L\sum_{n=1,2}s_x\prm{n}, \nonumber\\
&\mathcal{H}_d = \hbar g_d\tilde s_+\prm{1}\tilde s_-\prm{2} + \mathrm{H.c.}
 \end{align}
where $\tilde s_\pm\prm{n} \equiv s_y\prm{n} \pm i s_z\prm{n}$ for $n=1,2$. The parameter $g_d$ determines the coupling of the spins and describes a SWAP gate, in terms of the qubit operations.

We will employ the short notation $\Gr{A} \equiv \llangle A\vert B\rrangle_\omega$. The dependence on the operator $B$ is suppressed in this notation, as it is inessential for the following calculation. For concreteness we will assume $B=\tilde s_-\prm{1}$.

For the model (\ref{eq:toy_H}) we have from Eq.~(\ref{eq:Green_eom}) 
\begin{align}
\label{eq:Green_direct_cplng}
\Gr{\tilde s_+\prm{1}}(\omega - \omega_L) = 2g_d \Gr{s_x\prm{1}\tilde s_+\prm{2}} + \langle[\tilde s_+\prm{1},B]\rangle
\end{align}
with $ \langle[\tilde s_+\prm{1},B]\rangle = 2\langle s_x\prm{1}\rangle$. Equation (\ref{eq:Green_direct_cplng}) has the form that we will aim at obtaining where there is no direct  coupling between the spins.  

%\subsection{Coupling via a common vibrational mode}

The Hamiltonian of two electrons coupled to the same vibrational mode by the spin-orbit coupling has the form
\begin{align}
\label{eq:H_one_mode}
&H_{eev} = -\hbar \omega_L\sum_{n=1,2}s_x\prm{n} + \hbar\omega_0 a_0^\dagger a_0\nonumber\\
& +
\hbar (a_0+a_0^\dagger)\sum_{n=1,2} \zeta_ns_z\prm{n} 
\end{align}
Here $a_0,a_0^\dagger$ are the ladder operators of the mode, $\omega_0$ is its frequency, and $\zeta_{1,2}$ are the coupling parameters. We assume these parameters to be small compared to $\omega_0,\omega_L, |\omega_0-\omega_L|$. We further assume that the temperature is low so that $\langle a_0 a_0^\dagger\rangle \approx 1$. 

The equation of motion for the same Green function we considered earlier now reads
\begin{align}
\label{eq:first_iteration}
\Gr{\tilde s_+\prm{1}}(\omega - \omega_L) = i\zeta_1\Gr{(a_0+a_0^\dagger)s_x\prm{1}}+ \langle[\tilde s_+\prm{1},B]\rangle
\end{align}
We can now write similar equations for the Green function on the right-hand side. For example,
\begin{align}
\label{eq:second_iteration}
\Gr{a_0 s_x\prm{1}}(\omega - \omega_0) = -\frac{i}{2}\zeta_1\Gr{s_y\prm{1}}+\zeta_2\Gr{s_x\prm{1} s_z\prm{2}}
\end{align}
Here we have use that,  for low temperatures, $\Gr{a_0a_0^\dagger s_z\prm{2}s_x\prm{1}}\approx \Gr{s_z\prm{2}s_x\prm{1}}$.

The function $\Gr{\tilde s_+\prm{1}}$ is large for $\omega\approx \omega_L$. This is immediately seen from Eq.~(\ref{eq:first_iteration}). Such behavior can be understood from Eq.~(\ref{eq:Green_eom}), since for an isolated spin we have in the Heisenberg representation $\tilde s_+\propto \exp(-i\omega_Lt)$. Therefore in Eq.~(\ref{eq:second_iteration}) we can set $\omega = \omega_L$ and replace $\Gr{s_y\prm{1}}\approx (1/2)\Gr{\tilde s_+\prm{1}}$ and $\Gr{s_x\prm{1}s_z\prm{2}}\approx -(i/2) \Gr{s_x\prm{1} \tilde s_+\prm{2}}$. 

Making the aforementioned approximations in Eq.~(\ref{eq:second_iteration}) and in the similar equation for the function $\Gr{a_0^\dagger s_x\prm{1}}$ and substituting the result into Eq.~(\ref{eq:first_iteration}), we bring Eq.~(\ref{eq:first_iteration}) to exactly that same form as Eq.~(\ref{eq:Green_direct_cplng}), with the substitution
\begin{align}
\label{eq:renormalization_one}
&\omega_L \to \omega_L + \frac{1}{2}\zeta_1^2\omega_L/(\omega_L^2 - \omega_0^2),\nonumber\\
&g_d \to g_0, \qquad g_0=\frac{1}{2}\zeta_1\zeta_2\omega_0/(\omega_L^2 - \omega_0^2).
\end{align} 
The change of the Larmor frequency is the polaronic effect of the coupling to the vibrational mode, whereas $g_0$ is the parameter of the spin-spin coupling mediated by the coupling to this mode.

%%%%%%%%%%%%%%%%%%%%%%%%%%%%%%%%%%%%%%

%\bibliography{c:/Users/dykmanm/Dropbox/Aaa/BibTex/Zotero19}
%\bibliography{c:/Users/mark/Dropbox/Aaa/BibTex/Zotero19}


\end{document}